# Probabilistic Prime Factorization based on Virtually Connected Boltzmann Machine and Probabilistic Annealing

Hyundo Jung[1]*, Hyunjin Kim[2]*, Woojin Lee, Jinwoo Jeon, Yohan Choi, Taehyeong Park, and Chulwoo Kim[3]

Korea University, Seoul, Korea

1: guseh7274@gmail.com, 2: jamespul321@gmail.com, 3: ckim@korea.ac.kr

*: These authors contributed equally to this work

Probabilistic computing has been introduced to operate functional networks using a probabilistic bit (p-bit), generating 0 or 1 probabilistically from its electrical input. In contrast to quantum computers, probabilistic computing enables the operation of adiabatic algorithms even at room temperature, and is expected to broaden computational abilities in non-deterministic polynomial searching and learning problems. However, previous developments of probabilistic machines have focused on emulating the operation of quantum computers similarly, implementing every p-bit with large weight-sum matrix multiplication blocks[1] or requiring tens of times more p-bits than semiprime bits[2–3]. Furthermore, previous probabilistic machines adopted the graph model of quantum computers for updating the hardware connections, which further increased the number of sampling operations. Here we introduce a digitally accelerated prime factorization machine with a virtually connected Boltzmann machine and probabilistic annealing method, designed to reduce the complexity and number of sampling operations to below those of previous probabilistic factorization machines[1–3]. In 10-bit to 64-bit factorizations were performed to assess the effectiveness of the machine, and the machine offers $1.2 \times 10^8$ times improvement in the number of sampling operations compared with previous factorization machines, with a 22-fold smaller hardware resource. This work shows that probabilistic machines can be implemented in a cost-effective manner using a field-programmable gate array, and hence we suggest that probabilistic computers can be employed for solving various large NP searching problems in the near future.

**Introduction**

Deterministic computers have been developed to enhance computing power using nanoscale transistors. However, despite the increasing demand for solving combinatorial optimization problems, deterministic computers perform slow and inefficient search operations[4], and the process-scaling of transistors has been reaching its limit[5]. Quantum computers have been introduced to rapidly solve these non-deterministic polynomial (NP)-hard problems[6–8]. However, the number of qubits required for such cases is still larger than that of physically implemented qubits.

As a result of the above issues, probabilistic computing[13–24], which is realized by using magnetic tunnel junction (MTJ) or complementary metal–oxide–semiconductor (CMOS) technologies, has been proposed to cost-effectively replace quantum annealers[9–12], especially for factorization calculations[1–3]. These Ising machines have improved the factorization speed by four[2] and six[3] orders above central processing units. However, the hardware complexity and computation time of previous factorization machines sharply increase as the number of total probabilistic bits (p-bits) increases. Therefore, a novel algorithm and its hardware implementation are required for practical applications.

In this work, we propose a virtually connected Boltzmann machine (VCBM) showing less hardware complexity than the state-of-the-art probabilistic Ising machines. Also, we introduce a probabilistic annealing method for reducing the number of sampling operations of the Boltzmann machine. Furthermore, we employed three modulo operators (divide *X* and *Y* candidates by 3, 5, and 7) preceded in each case by a sieve to determine the best candidate (termed hereafter a "candidate sieve") and a decision block that consists of two modulo operators (divide the semiprime by best candidate), in order to further accelerate the performance of the factorization machine. To show its potential to the community, results for up to 64-bit factorization were produced using a field-programmable gate array (FPGA).

**Virtually connected Boltzmann machine**

The Boltzmann machine[25–27] is a type of constraint satisfaction network that can use its probabilistic dynamics to lead a system to ground states in combinatorial optimization problems. Since the weights on the connections between the spins of the Boltzmann machine are fixed after the problem is formulated, Boltzmann machines are frequently used for implementing Ising models in hardware. However, as shown in Fig. 1a, the hardware should be implemented with large spin-weight matrix calculation circuits due to the 3-body and 4-body terms of the Ising model[1]. On the other hand, when the Ising model is implemented with hidden bits to represent 3-body or 4-body terms[12], the large number of hidden bits increases the number of hardware connections to the fourth power of *N* bits, (see Methods for details). In 2022, restricted Boltzmann machine (RBM)[28]-based probabilistic computing was introduced[2], but the large number of hardware connections in the RBM still limits the application of these machines in large-scale general-purpose probabilistic computers. Moreover, previous Ising machines require the deployment of an additional deterministic computer that formulates the hardware connections of the Ising machine before solving problems[1–3, 13–24].

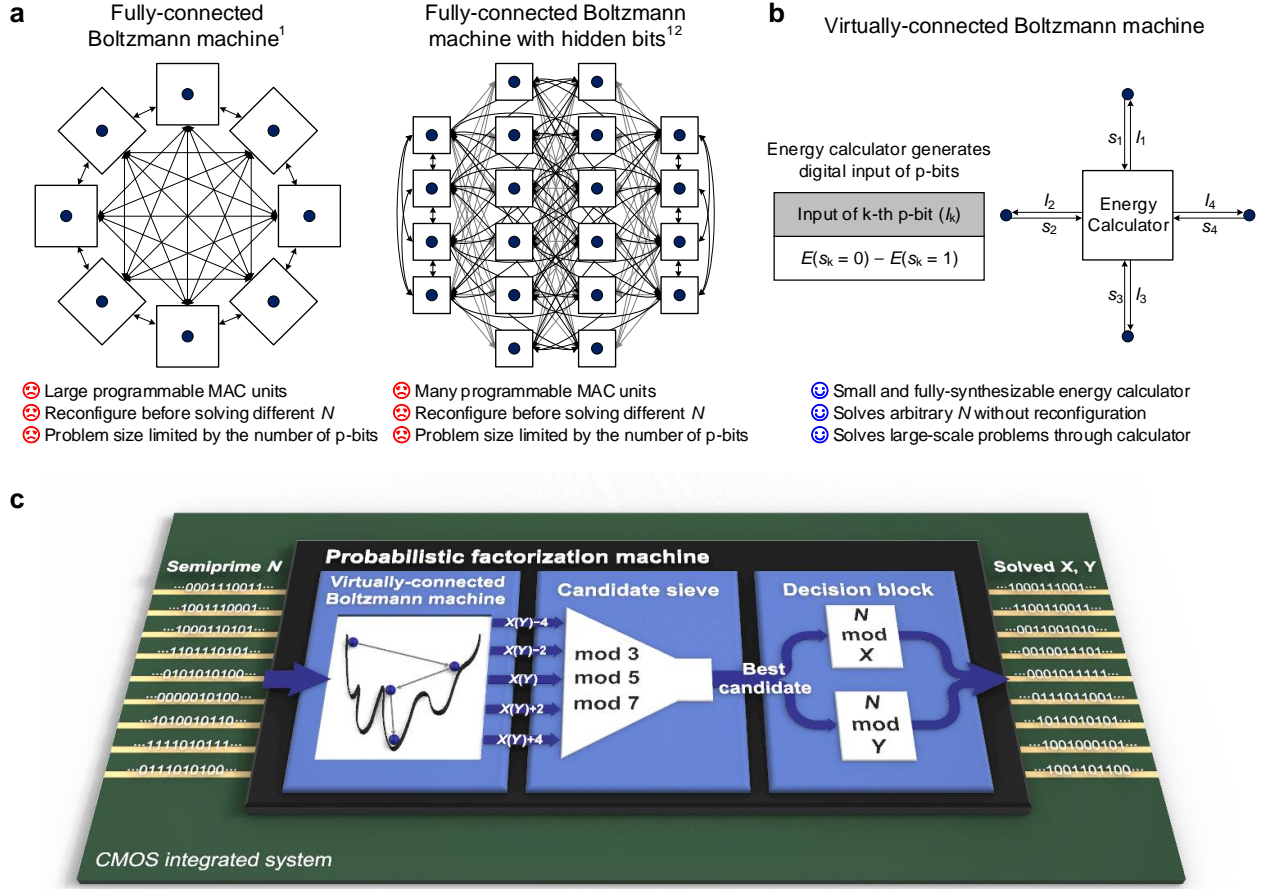

**Fig. 1 | Demonstration of the developed factorization machine. a, b,** Architectures of Boltzmann machine-based factorization machines that can perform up to 10-bit factorization. Circles represent p-bits, and squares represent the approximate hardware area of the digital logic. (**a**) In the graph model, p-bits of the general Boltzmann machine require a large area of spin-weight matrix multiplication logic[1]. This hardware cost of matrix multiplication logic of the general Boltzmann machine can be reduced by replacing the 3-body and 4-body terms with hidden p-bits[11–12]. However, hidden p-bits also increase hardware complexity. (**b**) Architecture of the VCBM. The digital input of p-bits can be calculated using the term $E(s_k=0) - E(s_k=1)$. Thus, the energy calculator generates p-bit inputs without spin-weight matrix multiplication. **c,** Architecture of our probabilistic factorization machine. In this work, we implemented a prototype of a 64-bit general-purpose factorization machine using an FPGA.

Figure 1b shows the high-level concept of the VCBM. In the general Boltzmann machine, the input value of the k-th p-bit ($I_k$) represents every connected interaction in the system to the visible p-bit. Thus, the computational power of CMOS digital circuits can be used to generate virtual connections of each visible p-bit. In our machine, the energy calculator is employed to calculate the digital input of each p-bit, replacing the need for a large and complex chain network composed of digital blocks or hidden p-bits. Since the VCBM uses an equivalent probability equation of input to p-bits as the general Boltzmann machine, the VCBM represents an all-to-all connected Boltzmann machine, which is suitable for solving large and complex combinatorial problems. Moreover, the energy calculator generates $I_k$ of each p-bit, enabling the factorization of an arbitrary semiprime $N$ without reconfiguring its weight connections between p-bits. Thus, our machine can perform from 10-bit to 64-bit factorizations of semiprimes continuously without additional hardware connection formulations. Detailed information on the derivation is provided in Methods.

Compared to the quantum annealers, we think that the major advantage of probabilistic annealers is their compatibility with digital CMOS circuits at room temperature. Thus, we propose a digitally accelerated probabilistic factorization machine architecture that operates synchronously using both digital logic and the Boltzmann machine, as shown in Fig. 1c. Considering that the Boltzmann machine frequently generates a high-quality output that is close to the global ground state, a set of nearby numbers of the output of the machine output also have high probability as an answer. Thus, we inserted a candidate sieve between the machine and the decision block to choose the candidate that is close to the output of the Boltzmann machine but cannot be divided by 3, 5, and 7 (best candidate) with small area consumption. Then, two modulo operators of the decision block, which are pipelined to operate in two cycles to reduce the critical-path delay, check whether $X$ or $Y$ is the factor of $N$. Therefore, the decision block determines the end of the factorization operation. For factorizing up to 64-bit numbers, the machine uses 31 p-bits that are implemented based on lookup tables (LUTs)[2,29] of the FPGA hardware. Detailed information on the FPGA implementation is provided in Methods.

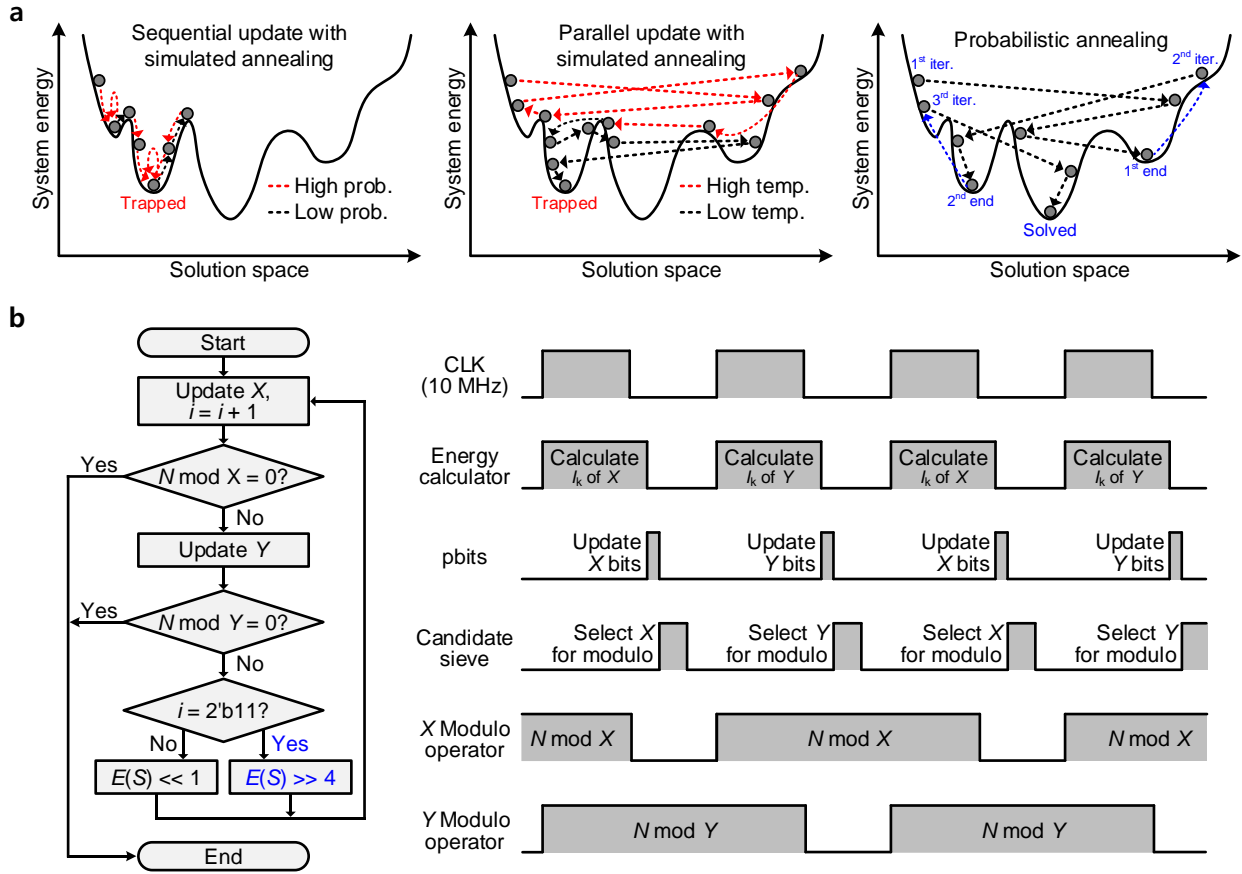

**Fig. 2 | Demonstration of probabilistic annealing. a,** Com-parison of state transitions in annealing processes. In general, sequential updating makes it difficult for the system to escape from a local minimum. In addition, previous machines performing parallel updating were designed to gradually decrease the temperature, causing the system to converge slowly and be trapped in a local minimum in its final state. The probabilistic annealing of the current work was designed to accelerate the system convergence to a minimum with parallel updating and dynamic SSPB control. Upon completion of 8 sampling operations (in which the number 8 is constant for every factorization operation in this work), the system restarts a searching iteration from a higher energy state until achieving the solution to the problem. **b,** Flowchart and operating sequence of the factorization machine. This machine is designed to employ a decision block ($X$ and $Y$ modulo operators) for determining the completion of factorization when $N$ mod $X$ or $N$ mod $Y$ becomes 0. The operation of the candidate sieve is conducted after the p-bit update (omitted in the flowchart for simplicity).

**Probabilistic annealing**
To further improve the factorization operation of the machine, we developed a probabilistic annealing process. The probabilistic annealing consists of performing parallel updates and controlling dynamic system-significant p-bit (SSPB). In previous works, the Ising machines reached the global minimum with sequential updating[1,14–24] or parallel updating[2–3,13], as shown in Fig. 2a. In the sequential update process, the system updates each bit to have lower energy after an update, forcing the system to climb the energy barrier through consecutive low-probability decisions. Thus, the sequential updating rather traps the system in a local minimum. On the other hand, parallel update machines were implemented for achieving shorter problem solving, using RBM[2], sparse Ising model[3], and stochastic cellular automata (SCA)[13,30]. However, the RBM machine and sparse Ising model require pre-training before each search operation. The SCA machine has been reported to solve max-cut problems with a level of quality equivalent to those of sequential update machines, but the machine converges the system energy through a slow logarithmic cooling along with penalty control, which rather fixes the system to a specific wrong solution in factorization problems. Moreover, previous parallel-updating machines did not achieve a computational complexity advantage over those using the sequential update method, which means they shortened the operation time at the expense of increasing the amount of hardware area.

Therefore, we introduce the probabilistic annealing to make the system rapidly converge to a minimum. In contrast to sequential updating, the simultaneous changing of p-bits would occasionally move system energy through an energy barrier with a single update. However, simultaneous updating of independent p-bits increases the divergence of the system energy. Considering an interaction between a system and p-bit, when the strength of the interaction is increased, the p-bit would be strongly fixed to 0 or 1. Otherwise, when the interaction is weak, the output of the p-bit would have an approximately 50% probability of being 0 or 1. If a p-bit has an adequate amount of interaction with the system, the p-bit would become an SSPB and be highly likely to induce the system to decrease its system energy. Therefore, constraining the number of SSPBs to a low value and dynamically changing SSPBs to other p-bits after the update are required to have a system converge rapidly to a ground state.

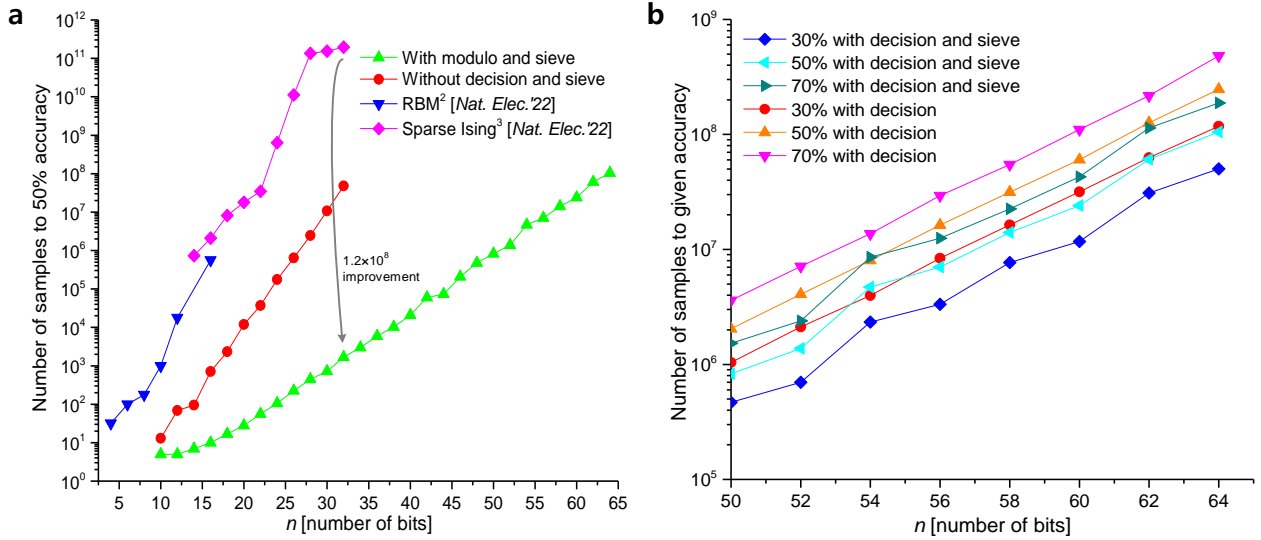

**Fig. 3 | Performance of the machine for factorization compared to previous parallel update factorization machines. a,** Results of 4-bit to 64-bit factorization measurements of the previous works[2–3] and the current work are shown. For a fair comparison between annealing schemes, performances without a candidate sieve and decision block are also shown. We repeated the experiment 1,000 times, and the number of sampling operations of the conventional sparse Ising machine[3] is calculated using its 100% time-to-solution results. As shown in the graph, the probabilistic annealing achieved a performance improvement of up to $1.4 \times 10^4$ times at 30-bit factorization, and the digitally accelerated architecture reduced the number of sampling operations by $1.2 \times 10^8$ times than the previous work[3] at 32-bit factorization. **b,** Detailed factorization results from 50-bit to 64-bit with decision block are shown. Using the candidate sieve reduced the number of sampling operations at 50% accuracy by up to 66% (66% reduced at 52-bit factorization) with a small hardware cost. Detailed information about the sampling frequencies is provided in Methods.

We implemented dynamic SSPB control with a shift register in the energy calculator along with the cost function[36] $E(S) = E_0(XY - N)^2$. In this work, the coefficient of the cost function ($E_0$) is $2^{3-2n}$, where $n$ represents the number of digital bits of $N$. Since $E(S)$ is a most-significant-bit (MSB)-dominant function, $E(S)$ naturally constrains a small number of p-bits to become SSPBs, depending on the current energy state of the system. Then, by using the shift register to left-shift $E(S)$, the system changes the SSPBs from MSB to a less significant bit of $X$ and $Y$ during the factorization operation. Therefore, the probabilistic annealing periodically changes SSPBs of $X$ and $Y$ from the MSBs to the least significant bits (LSBs), forcing the system to converge to a minimum rapidly.

Figure 2b shows the operating algorithm of the factorization machine. Considering that $N$ is factored into two independent prime numbers, the machine updates $X$ and $Y$, targeting to converge $X$ and $Y$ to different prime numbers. After each update, and the candidate sieve conducts the modulo operation by 3, 5, and 7 with four candidates nearby $X$ ($X - 2$, $X$, $X + 2$, $X + 4$) or $Y$ ($Y - 2$, $Y$, $Y + 2$, $Y + 4$). When $X$ is updated, the candidate sieve simultaneously divides $X - 2$, $X$, $X + 2$, $X + 4$ by 3, 5, and 7, and choose the best candidate that is not divisible from the sequence of $X$, $X + 2$, $X - 2$, and $X + 4$. If all of these candidates are divisible by 3, 5, or 7, the candidate sieve selects $X - 4$ as the best candidate because $X - 4$ is naturally indivisible by 3, 5, and 7. Therefore, the machine only conducts decision operations on a candidate that cannot be divided by 3, 5, or 7. After a consecutive $X$ and $Y$ update, the machine operates the left-shift of the cost function $E(S)$ for conducting the probabilistic annealing. Then, upon completing 8 sampling periods (4 samplings per $X$ and $Y$ each), the machine resets the bit-shifted $E(S)$ to the original state, making the system have a high energy state again. Thus, our probabilistic annealing process is not designed to find the answer with a single and long iteration as in previous works, but rather to repeat less-energy-barrier-constrained searching iterations toward a minimum with 8 sampling periods. In contrast to previously developed factorization machines requiring static random-access memory (SRAM) to store weights and biases[2] or use of the MATLAB program to multiply input weights and inverse temperature online[3], the probabilistic annealing does not have to optimize the weight values of the factorization machine through pre-training and does not have to calculate the complicated temperature equation of the system. Therefore, this work fully conducts operations with synthesized digital gates in an FPGA, achieving a dramatic reduction in hardware complexity.

| Platform | Machine architecture | Annealing method | # of LUTs | # of spins | Bits factorized |
|---|---|---|---|---|---|
| D-wave 2000Q[11] | Ising | Quantum | - | 1,803 | 18 |
| Stochastic MTJ[1] | General Boltzmann | - | - | 8 | 10 |
| FPGA[2] | Restricted Boltzmann | - | 1,182,240 | 680 | 16 |
| FPGA[3] | Sparse Ising | Simulated | 1,182,240 | 2,128 | 32 |
| FPGA (this work) | Virtually connected Boltzmann | Probabilistic | 53,200 | 31 | 64 |

**Table 1 | Comparison of Ising-model-based factorization machines.** Here we compare the hardware performances of the state-of-the-art Ising-model-based factorization machines. The RBM work[2] used quantization retraining for performance enhancement. Compared to the FPGA (Xilinx Virtex UltraScale+ VU9P) used in recent works[2–3], our current work employed a more cost-effective FPGA (Xilinx Artix-7) owing to its small hardware requirements. As shown in the table, our machine factorizes the largest semiprime number with the lowest hardware cost and the smallest number of spins per unit number of semiprime bits.

**Experiment**

Figure 3a shows the results of the measurements in the current work compared with those of previous works. For performance comparison between annealing methods, we also measured the factorization machine operating without both candidate sieve and modulo operator that finishes the factorization only when $XY = N$ as the previous works[1–3]. The measurement results show that the probabilistic annealing reduces the number of factorization operations by up to $1.4 \times 10^4$ times, compared to previous parallel update machines[2–3]. Moreover, when we employed the candidate sieve and decision block for the factorization operations, there was an $1.2 \times 10^8$ times reduction in the number of sampling operations at 32-bit factorization compared to the previous work[3]. Figure 3b shows a graph of the detailed performance results using the candidate sieve and decision block. As seen in this graph, the use of the candidate sieve decreased the number of sampling operations by up to 66% at 50% accuracy; the number of samples at 50% accuracy represents the measured number of samples that can factorize 500 out of 1,000 experiments. Since the critical path delay of our machine is the operation of the decision block, adding the candidate sieve increases only 1.56 ns of additional operation time. The sieve costs 1,466 (3.82% of used) lookup tables (LUTs), thus, we demonstrate that the candidate sieve increased factorization performance with a low amount of hardware consumption. Time-domain measurement results of our factorization machine are shown in Extended Fig. 1.

Table 1 summarizes the performances of state-of-the-art Ising-model-based factorization machines. Due to the deployment of the VCBM, our work requires the fewest p-bits and is implemented without the multiply-accumulate (MAC) unit for weight and spin multiplications, which consume the majority of LUTs in previous probabilistic machines[2–3]. Therefore, compared to recent works[2–3], we employed a relatively cost-effective FPGA (Xilinx Artix-7) with a smaller amount of programmable logic (PL). Also, contrary to the previous works[2–3] that require more complex Ising machine hardware for parallel updating, the probabilistic annealing enables parallel updating with the fully-connected VCBM. Moreover, due to the high scalability and rapid factorization performance of our machine, we carried out up to 64-bit factorization, which is the highest in the table.

Figure 4 shows the test setup and measurement results with four independent FPGAs. In 2021, a multi-chip architecture based on simulated bifurcation[18] reported up to $1.89\times$ of computation time reduction by using two chips on a single problem. However, the computation time was reduced by a factor of only 3.32 (i.e., less than 4.00) with four chips, implying that the simulated bifurcation would result in per-chip performance degradation in large-scale multi-chip processor applications. On the contrary, since our VCBM represents the fully-connected Boltzmann machine, the average factorization performance is improved by the number of parallel-connected chips. When factorizing with a decision block, our machine improves the average factorization performance by factors of $2.01\times$, $3.05\times$, and $3.98\times$ on average when using two, three, and four chips, respectively. Furthermore, our machine with a candidate sieve improves the average factorization performance by factors of $2.07\times$, $3.17\times$, and $4.22\times$ when using two, three, and four chips, respectively. Unlike previous annealing processors[18,33], our machines reduce the operation time without data interaction between each other, thus, the hardware resources can be easily shared and reconfigured according to the size of the problem to be solved. Further detailed measurement results are shown in Extended Fig. 2 and 3.

**Discussion and outlook**

In this work, we have demonstrated a probabilistic factorization machine based on a VCBM and probabilistic annealing scheme. To overcome the hardware complexity of previous Ising machines, we developed the VCBM. The VCBM was designed to update its p-bits using an energy calculator, eliminating the need for complex matrix multiplications. Moreover, unlike previous Ising machines, our machine was designed to factor various semiprime numbers without reconfiguring hardware connections, enabling the continuous operation of the annealer. As a result, we factorized the highest number with the smallest number of spins and the lowest hardware cost among the state-of-the-art Ising-model-based factorization machines. Also, we proposed the probabilistic annealing method to enable parallel updating with rapid convergence to the global ground state. Thus, the machine achieved a factorization performance of up to $1.4 \times 10^4$ times faster than previous annealing methods.

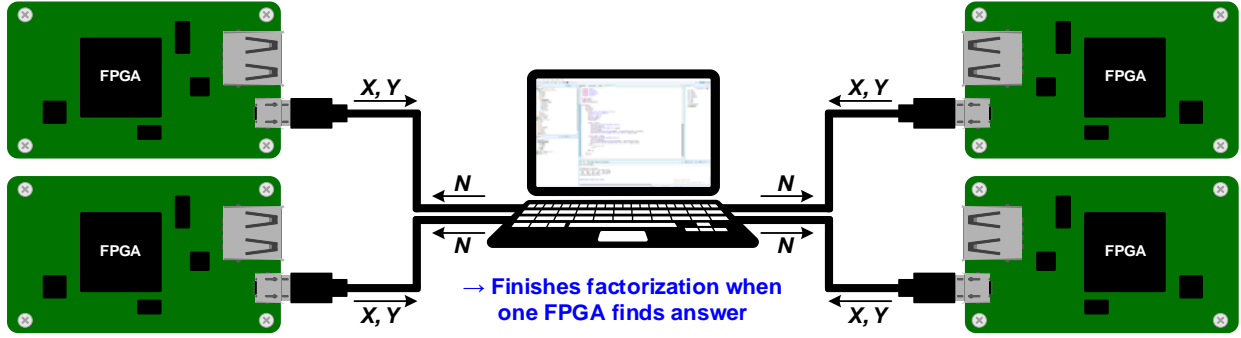

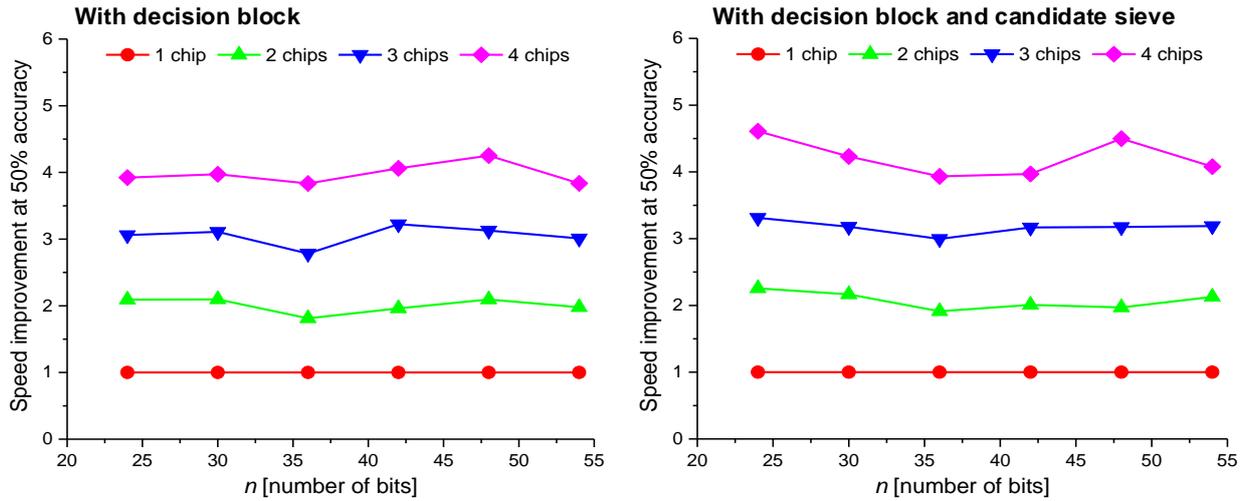

**Fig. 4 | Performance of the machine for multi-chip computing.** Measurement environment and measured results of multi-chip factorizations are shown. Four equivalent FPGA boards were connected to a computer and started operation together to factor the equivalent semiprime $N$. The factorization was finished immediately when one of the FPGA boards factorized $N$. The measurements were repeated 1,000 times, and factorization speed improvement (y-axis) represents the normalized number of samples of multi-chip architecture compared to that of the one-chip architecture. Compared to single-chip computing, multi-chip computing with 2, 3, and 4 machines achieved approximately 2.01×, 3.05×, and 3.98× reductions in the number of sampling operations at 50% accuracy (500 solved experiments) with decision block. When the machine with a candidate sieve is employed, the average factorization performance is improved by 2.07×, 3.17×, and 4.22× when using two, three, and four chips, respectively.

Furthermore, we have shown that using digital logic (candidate sieve and decision block) can be a key to enhancing the performance of CMOS-based Ising machines. As a result, the number of sampling operations was decreased $1.2 \times 10^8$-fold at 32-bit factorization compared to the previous factorization machine[3]. Although the prototype of this work was implemented in an FPGA, we expect that our machine can be implemented using MTJ-based p-bits[34–35] and a hardware-optimized energy calculator in a single package, allowing solving larger semiprime numbers in a short time with lower computing power.

## Methods

**Derivation of the virtually connected Boltzmann machine**

In general, in the Ising model[37–38], the energy function $E(S)$ of the system with spin vector $S$ is

$$E(S) = -(\sum_{i<j} w_{ij} s_i s_j + \sum_i h_i s_i), \quad (1)$$

where $s_i$ and $s_j$ are state of the i-th and j-th spin, $w_{ij}$ is the weight of the edge between the i-th and j-th spins, and $h_i$ is a bias term for the i-th spin. Consider defining the probability $p$ of the system with status $S$ as[25–27, 39]

$$p(S) = \frac{\exp(-E(S))}{Z}, \quad (2)$$

where $Z$ is the sum of all the cases of system energy and given using the equation

$$Z = \sum_S \exp(-E(S)). \quad (3)$$

Then, in a Boltzmann machine, the lower the energy of state $S$, the higher the probability appears, which means that the system more likely moves to a lower energy state after transitions. Therefore, the energy function is set to have the minimum energy state at the solution state, the Boltzmann machine would move the system energy toward the solution state. Consider the conditional probability $p(s_k = 1 \mid S)$ as the probability that the k-th spin will be updated to 1 in the next transition at state $S$,

$$p(s_k = 1 \mid S) = \frac{p(s_1, s_2, \ldots, s_{k-1}, 1, s_{k+1}, s_{k+2}, \ldots)}{p(s_1, s_2, \ldots, s_{k-1}, 0, s_{k+1}, s_{k+2}, \ldots) + p(s_1, s_2, \ldots, s_{k-1}, 1, s_{k+1}, s_{k+2}, \ldots)}$$

$$= \frac{\exp(-E(s_1, \ldots, s_{k-1}, 1, s_{k+1}, s_{k+2}, \ldots))/Z}{\exp(-E(s_1, \ldots, s_{k-1}, 0, s_{k+1}, s_{k+2}, \ldots))/Z + \exp(-E(s_1, \ldots, s_{k-1}, 1, s_{k+1}, s_{k+2}, \ldots))/Z}$$

$$= \frac{1}{1 + \exp(E(s_1, \ldots, s_{k-1}, 1, s_{k+1}, s_{k+2}, \ldots) - E(s_1, \ldots, s_{k-1}, 0, s_{k+1}, s_{k+2}, \ldots))}. \quad (4)$$

Thus, by defining $E(s_k = 1)$ and $E(s_k = 0)$ as, respectively,

$$E(s_k = 1) = E(s_1, \ldots, s_{k-1}, 1, s_{k+1}, s_{k+2}, \ldots), \quad (5)$$

$$E(s_k = 0) = E(s_1, \ldots, s_{k-1}, 0, s_{k+1}, s_{k+2}, \ldots), \quad (6)$$

and by defining $I_k = E(s_k = 0) - E(s_k = 1)$, the conditional probability becomes

$$p(s_k = 1 \mid S) = \frac{1}{1 + \exp(-I_k)}. \quad (7)$$

Therefore, if p-bits follow the probability of a sigmoid function to the input $I_k$ value, then the machine would follow the Boltzmann distribution[25–27, 39].

In our work, we utilized the energy function $E(S) = E_0 (XY - N)^2$, and thus the input value of the k-th bit ($I_k$) for updating $X$ can be simplified as,

$$I_k = E(s_k = 0) - E(s_k = 1)$$
$$= E_0 (X_{k,0} Y - N)^2 - E_0 (X_{k,1} Y - N)^2$$
$$= E_0 (X_{k,0} - X_{k,1}) ((X_{k,0} + X_{k,1}) Y - 2N) Y$$
$$= 2^{2+k-2n} (2N - (X_{k,0} + X_{k,1}) Y) Y$$
$$= 2^{3+k-2n} (N - XY) Y \pm 2^{1+2k-2n} Y^2 \quad (8)$$

with $n$ representing the number of bits of $N$, coefficient $E_0$ being $2^{3-2n}$, and $X_{k,0}$ and $X_{k,1}$ representing the $X$ values with 0 and 1 at k-th bit, respectively. Therefore, the energy calculator calculates $(N - XY)Y$ and $Y^2$ only once, and multiplies them by $2^{3+k-2n}$ and $2^{1+2k-2n}$ simply using the shift register. Then, if the k-th bit of $X$ ($X_k$) is 1, the calculator adds these two terms, otherwise, the calculator subtracts them. Therefore, the VCBM represents a fully-connected Boltzmann machine without the need to carry out weight-spin multiplication and accumulation operations. Due to the generality of the derivation of the VCBM, the machine can also be applied to other NP-hard problems that the Boltzmann machine can represent.

In VCBM, only $(n / 2 - 2)$ visible p-bits are required since $X$ and $Y$ are updated in turn. In addition, considering the hardware cost of the $(N - XY)Y$ multiplier, the computational complexity of the VCBM is determined to be $O(n^3)$, where $n$ is the number of digital bits of semiprime $N$.

**Hardware cost of previous Boltzmann machines**

In hardware-implemented Ising machines, inputs of p-bits are generated by adding a bias term to the sum of the product of the weight and output of connected spins[37]. However, as the size of the problem is increased, the number of required p-bits increases, and accordingly, the number of p-bit connections dramatically increases. Since the computational complexity of digital logic shows the effectiveness of the hardware, we also derived the hardware costs of previous fully connected Boltzmann machines. For a fair comparison, the hardware costs were considered with the machine that has the energy function $E(S) = E_0 (XY - N)^2$ and factorizes $n$-bit semiprimes with $(n/2 - 1)$-bit $X$ and $(n/2 - 1)$-bit $Y$.

In the general Ising model[37], the energy function $E(S)$ can be expressed as, $(\sum_i x_i 2^i \cdot \sum_j y_j 2^j - N)^2$, where $x_i$ and $y_j$ are the i-th and j-th binary bits of $X$ and $Y$, respectively[12, 39]. Then, this function can be classified into five different terms:[1] an 1-body constant bias term, a 2-body $x_i \cdot y_j$ term, a 3-body $x_i \cdot x_k \cdot y_j$ and $x_i \cdot y_j \cdot y_k$ terms, and a 4-body $x_i \cdot x_k \cdot y_j \cdot y_l$ term. However, for implementing the hardware with simple weight-spin matrix operators, the general Ising model should be expressed without 3-body and 4-body terms[11–12]. Thus, $3\binom{n/2-1}{2}$ hidden p-bits are produced for eliminating 3-body terms, and $(n/2 - 1)\binom{n/2-1}{2}$ hidden p-bits are required to eliminate 4-body terms, as shown in Fig. 2b. Therefore, $(n^3 - 4n^2 - 4n + 16) / 16$ hidden p-bits are required for one additional visible p-bit in the general Ising model. Assuming the use of fixed-point weight bits by the machine, 2-body terms of each p-bit require summation operations of the weight-spin matrix proportional to $n^3$. In conclusion, based on the total number of p-bits being proportional to $n^4$, the computational complexity of the machine is determined to be $O(n^7)$.

The Ising model can also be implemented by using 3-body and 4-body terms to express the energy function without hidden p-bits[1], as shown in Fig. 2a. Then, the Ising machine requires $n - 2$ visible p-bits for representing $X$ and $Y$.

However, $(n/2 – 1)\binom{n/2-1}{2}$ 4-body terms are required to be summed for calculating $I_k$. As a result, each p-bit requires calculations proportional to $n^3$, making a computational complexity of $O(n^4)$.

Moreover, the hardware complexity of previous probabilistic machines dramatically increases during the digital implementation of the annealing process. Due to the precision of weights determining the hardware accuracy of the Ising machine, precise time-dependent weight calculation logic inevitably increases hardware area in large semiprime factorizations. Also, these required weight values vary for factoring different semiprimes[11–12], making it difficult to implement solely in FPGA hardware. Therefore, the previous work[3] implemented a weight and inverse temperature multiplication by co-running the MATLAB program online. However, our machine was designed without these complex weight and temperature multiplications, enabling the FPGA to conduct factorization operations itself.

**FPGA implementation**
We programmed the factorization machine using Verilog-HDL, and coded and simulated using the Xilinx Vivado 2020.2 program. To achieve a shorter operation time, our factorization machine was programmed to conduct sampling operations of every p-bit in one clock cycle. Therefore, the physical computation time of our work can be derived by multiplying the number of sampling operations by the clock frequency (5 MHz for 64-bit).

For implementing FPGA hardware, p-bits are modeled with LUT-based on sigmoid function and a 48-bit pseudo-random linear-feedback shift register (LFSR)[2,29]. To accurately show the performance of our machine, we did not quantize the integer multiplication operations of the energy calculator. After the integer calculation is finished, the energy calculator generated 8-bit $I_k$ as a result of the shift register, which consisted of 1 sign bit, 3 integer bits and 4 fractional bits. Thus, the input of the sigmoidal activation function ($I_k$) ranges from -8 to 8 in real numbers, and the sigmoid function generates 16-bit output with $I_k$ input. Thus, the digital comparator compared this 16-bit output of the sigmoid function to the 16-bit output of the LFSR. If the output of the activation was higher, the p-bit was assigned 1 in the next update, otherwise, it was assigned 0 in the next update. For performing up to 64-bit factorizations, we used a 48-bit LFSR per p-bit that generates 16 pseudo-random bits per clock cycle.

Our 64-bit factorization machine used 38,086 (71.59% of total) LUT resources. However, since the decision block (X and Y modulators) required 4,455 (8.38% of total) LUT resources and the candidate sieve required 1,455 (2.73% of total) LUT resources, these digital circuits accelerated the factorization performance with low hardware cost. We employed a 5 MHz clock for performing up to 64-bit factorizations, and the operation of the candidate sieve generated a critical-path delay. Also, our machine can dynamically operate with a faster clock for a smaller target semiprime $N$, by disabling the factorizing operation of unnecessary upper bits of $X$ and $Y$.

After a single synthesis of a 64-bit factorization machine, we measured the sampling times by changing the number $N$ and the seed number for LFSRs. For practical purposes, our machine generated seed numbers of LFSRs internally, using a single 32-bit seed number input from the computer. We designed an Advanced eXtensible Interface (AXI) IP block to transmit two input data (a 64-bit subprime $N$ and 32-bit seed number) and a System Integrated Logic Analyzer (System ILA) IP block to read three output data (31-bit $X$, 31-bit $Y$, and 64-bit operation time). The processor system blocks were operated with a 25 MHz clock, while the VCBM factorization IP block and other blocks were operated with a 5 MHz clock. The Vitis 2020.2 program was employed to process the read and write data of the FPGA via the USB JTAG interface.

**Model cross-validation**
We implemented and tested the equivalent Boltzmann machine using MATLAB and Simulink to determine the accuracy of the measurements. From 10-bit to 32-bit semiprime numbers were tested 1,000 times each, and a comparison with the FPGA measurement results is shown in Extended Data Fig. 4. We used a Mersenne Twister(MT)-based uniform random generator model (MT19937ar)[41], which has a period of a sequence of $2^{19937} − 1$, instead of the 48-bit LFSR in p-bit. Since the FPGA hardware is programmed to initialize the $X$ and $Y$ to start with the deterministic value, the factorization at low-number bits require more samples than those of the MATLAB simulation results. However, when factorizing above 24-bit, the FPGA experiments involved only a ±7% difference in the number of samples. Thus, we verified that the factorization machine is properly implemented in hardware with sufficient randomness.

**Data availability**
The data to reproduce the figures within this work is available from the corresponding authors upon reasonable request.

**Code availability**
The code within this work is available from the corresponding authors upon reasonable request.

**Acknowledgements** This work was supported by Samsung Research Funding & Incubation Center of Samsung Electronics under Project Number SRFC-IT2101-03.

**Author contributions** H.J., H.K., W.L., J.J., Y.C., T.K., and C.K. contributed to building the experimental setup, performed the measurements, and analyzed the data. H.J., H.K., and C.K. performed the theoretical analysis and co-wrote the manuscript. All work was supervised by C.K. and all authors discussed the results and contributed to the manuscript.

**Competing interests** The authors declare no competing interests.

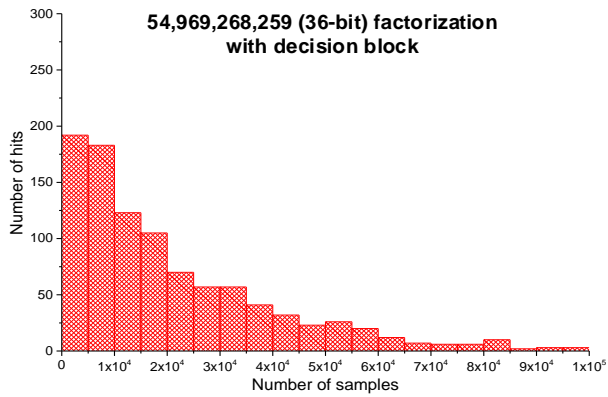
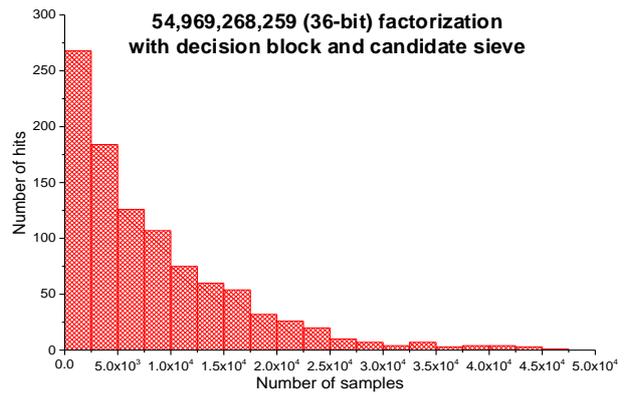
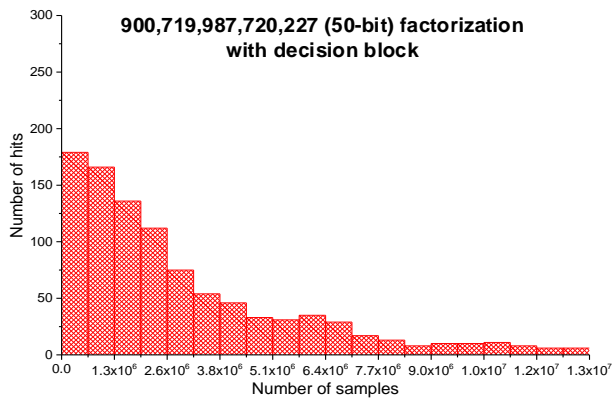
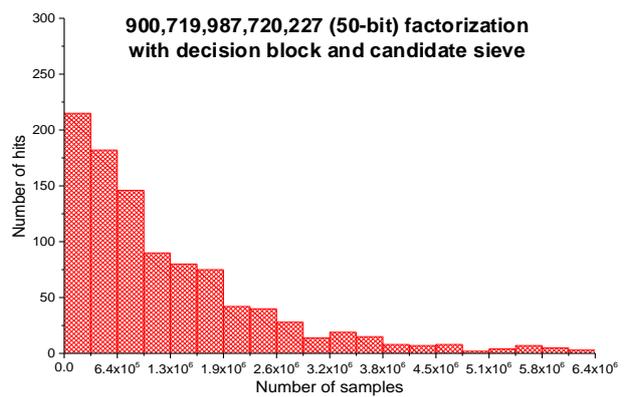
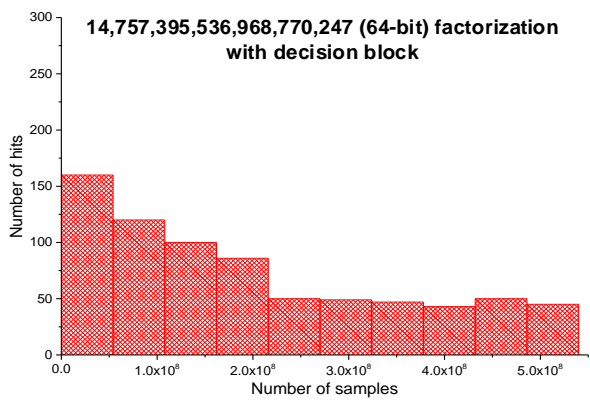
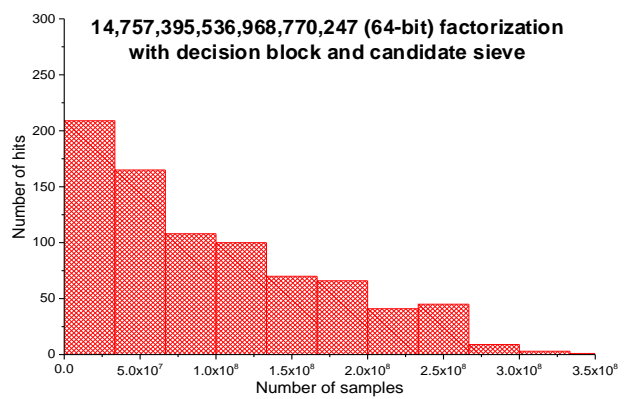

**Extended Data Fig. 1 | Time-domain histogram of factorizations.** The time-domain factorization experiment results are shown for further analysis. Our machine was experimented with the candidate sieve and decision block, and the histograms show that most experiments finished earlier than the middle point of x-axis, due to the Boltzmann distribution.

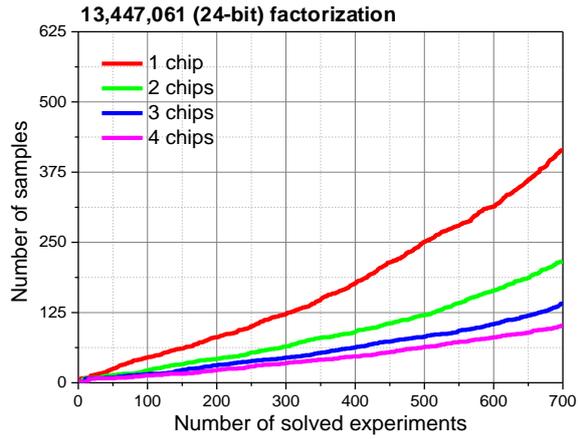
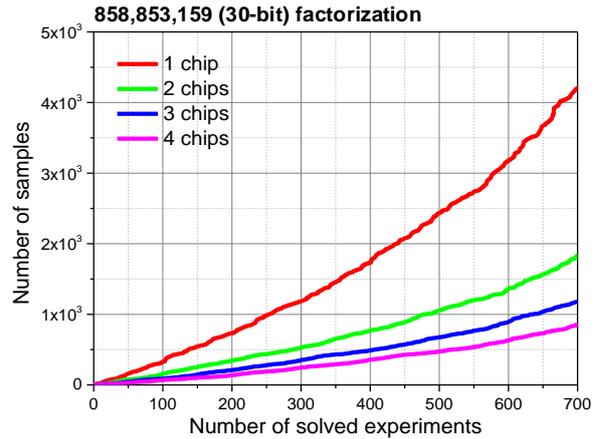
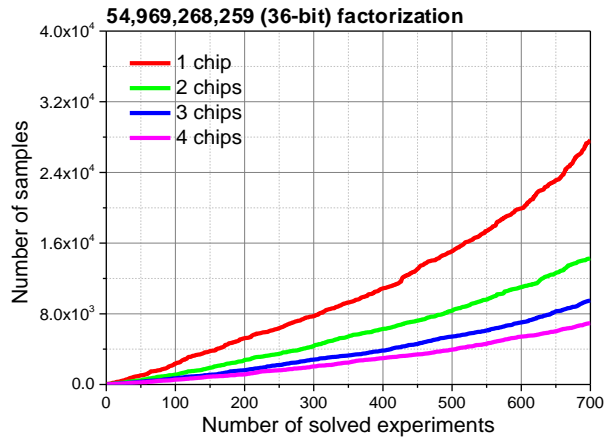
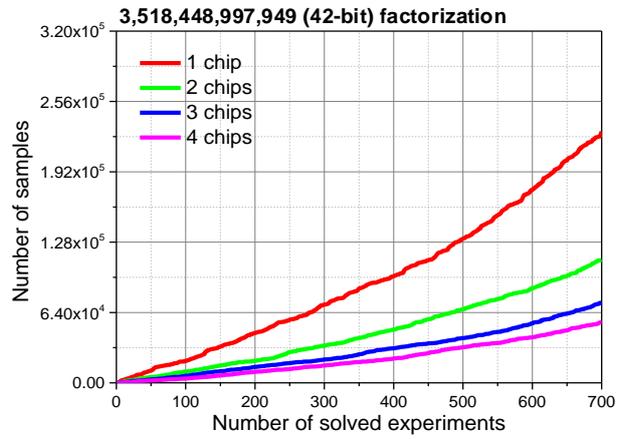
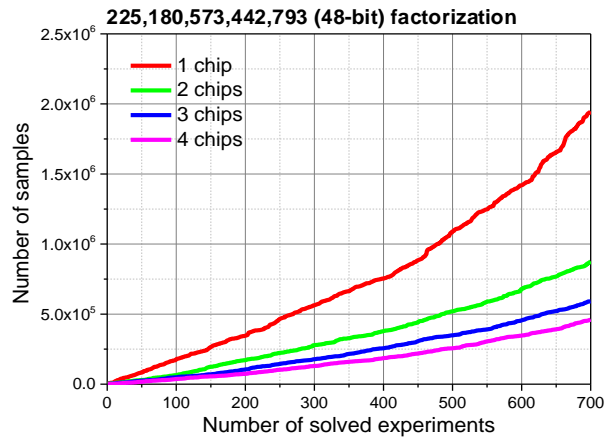
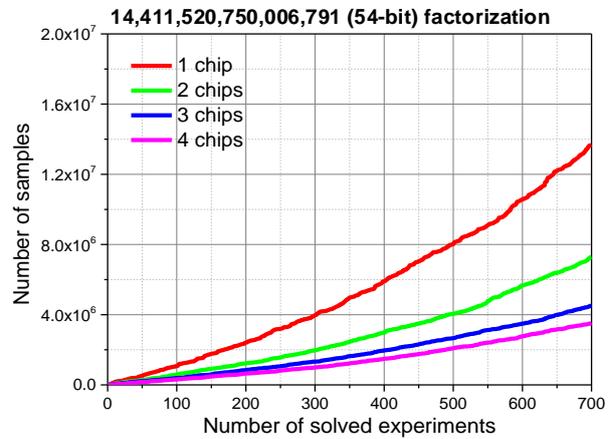

**Extended Data Fig. 2 | Number of required samples in multi-chip computations with the decision block and without the candidate sieve.** The cumulative factorization results are shown for analyzing the performance of the multi-chip computation. Our machine was found to improve the factorization performance by approximately 2×, 3×, and 4× when using two, three, and four chips, respectively.

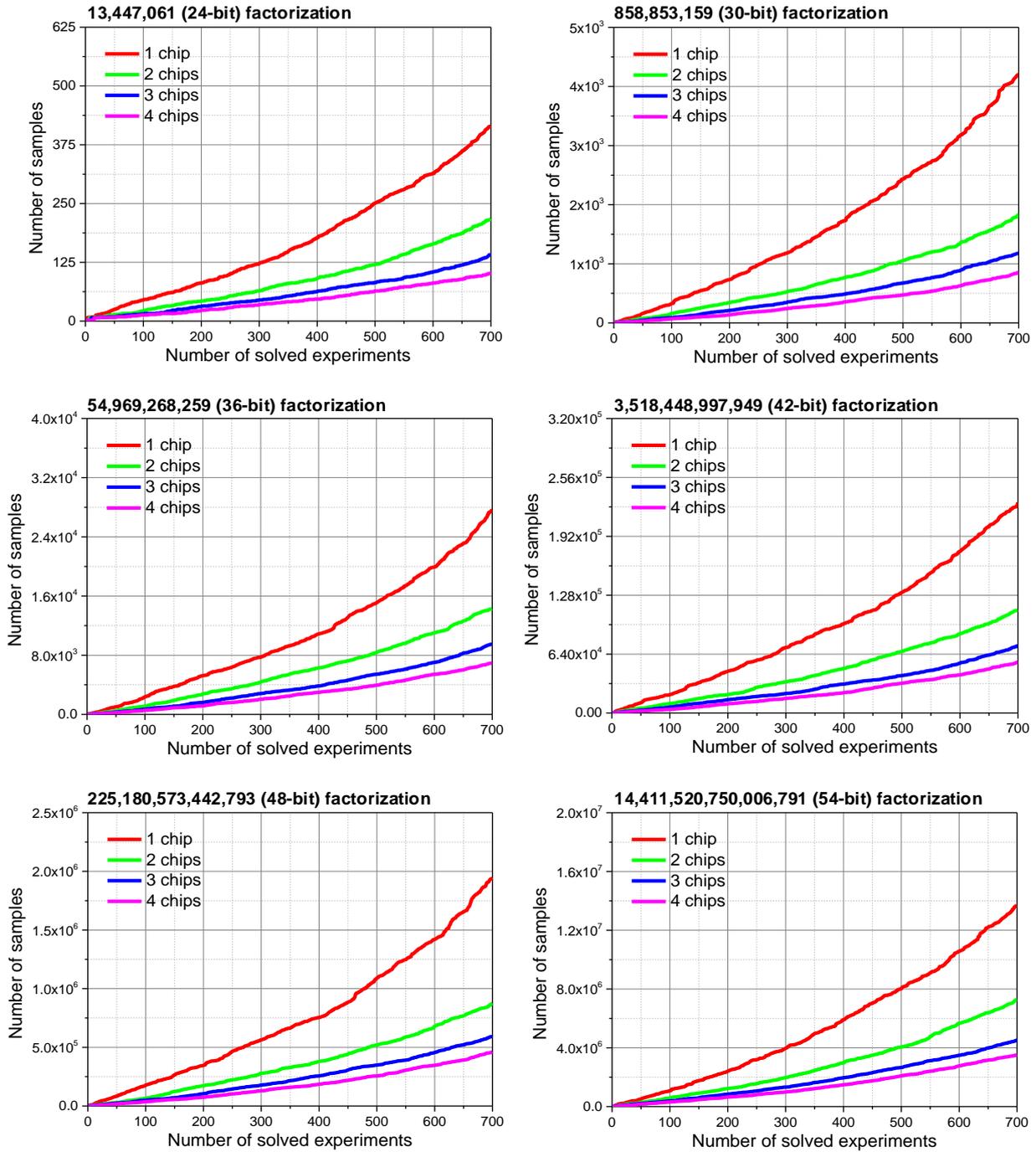

**Extended Data Fig. 3 | Number of required samples in multi-chip computations with the candidate sieve and decision block.** Additional cumulative factorization results are shown for analyzing the performance of the multi-chip computation with the candidate sieve and decision block. Our machine with candidate sieve also improves the factorization performance by approximately 2×, 3×, and 4× when using two, three, and four chips, respectively.

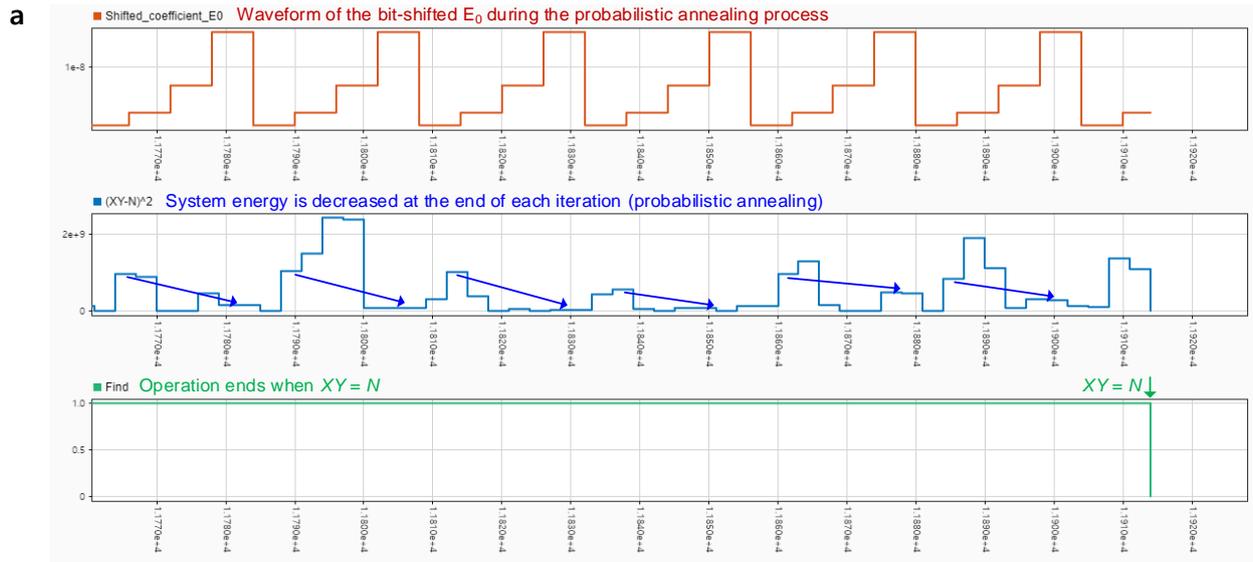

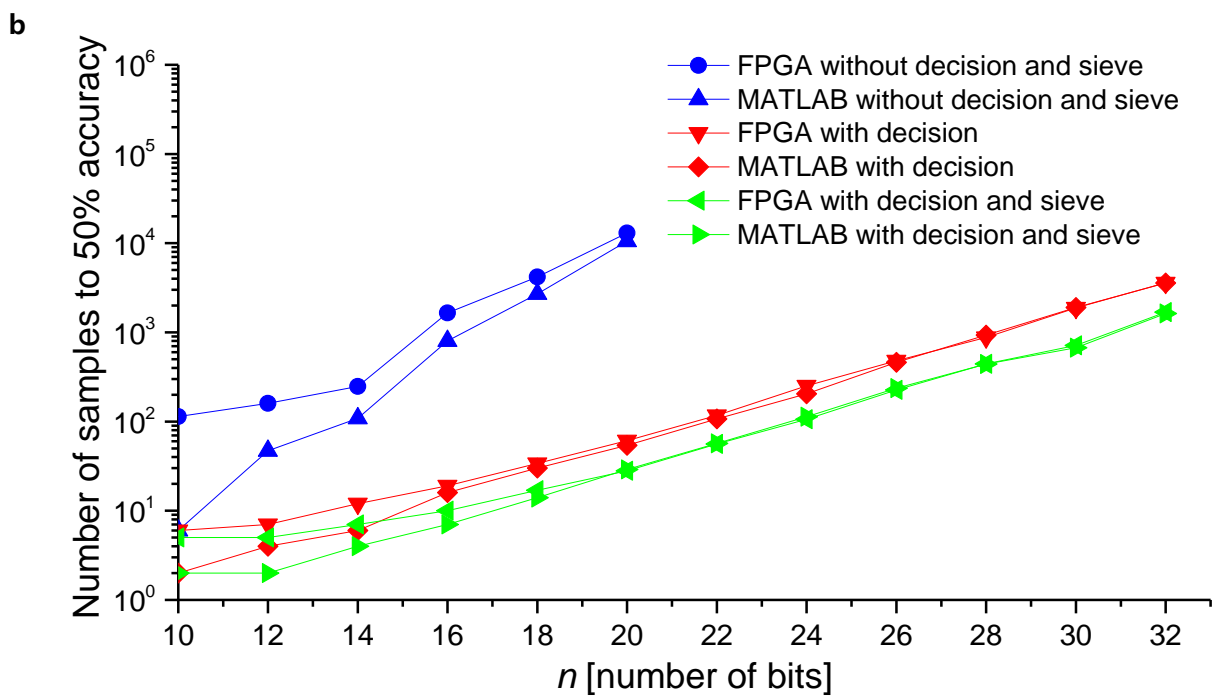

**Extended Data Fig. 4 | MATLAB Simulink simulation results compared to those of the FPGA experiments. a,** Simulated waveform of 16-bit factorization operation without decision block and candidate sieve using MATLAB Simulink. The Simulink model was designed to finish the factorization after the decision block determines the end of the operation. **b,** Results of 10-bit to 32-bit factorizations with the modulo operator when using MATLAB Simulink simulations compared to those obtained from FPGA experiments. For more exact model cross-validation, we tested semiprimes equivalent to those used in FPGA experiments.